\documentclass[prb,a4paper,twocolumn,superscriptaddress,showpacs,preprintnumbers,citeautoscript]{revtex4}
\usepackage[T1]{fontenc}
\usepackage[latin1]{inputenc}
\usepackage{graphicx,color,booktabs,microtype,afterpage,sidecap}
\usepackage[charter]{mathdesign}
\usepackage[colorlinks,plainpages=false,linkcolor=black,urlcolor=blue,citecolor=black,pdfpagemode=UseNone,pdfstartview=FitBH]{hyperref}

\pdfoutput=1

\makeatletter
\renewcommand{\@evenfoot}{\hfill\raisebox{-3em}{\bf\thepage}\hfill}
\renewcommand{\@oddfoot}{\hfill\raisebox{-3em}{\bf\thepage}\hfill}
\makeatother

\begin{document}

\title{Magnetic properties of PdAs$_{2}$O$_{6}$: a dilute spin system with
an unusually high N\'eel temperature}
\author{M. Reehuis}
\affiliation{Helmholtz-Zentrum Berlin f\"ur Materialien und Energie, D-14109 Berlin,
Germany}
\author{T. Saha-Dasgupta}
\affiliation{S. N. Bose National Center for Basic Sciences, Salt Lake, Kolkata 700098,
India}
\author{D. Orosel}
\affiliation{Max Planck Institute for Solid State Research, D-70569 Stuttgart, Germany}
\author{J. Nuss}
\affiliation{Max Planck Institute for Solid State Research, D-70569 Stuttgart, Germany}
\author{B. Rahaman}
\thanks{Present address: Aliah University, DN-41, Sector-V, Salt Lake,
Kolkata 700091, India}
\affiliation{S. N. Bose National Center for Basic Sciences, Salt Lake, Kolkata 700098,
India}
\author{B. Keimer}
\affiliation{Max Planck Institute for Solid State Research, D-70569 Stuttgart, Germany}
\author{O. K. Andersen}
\affiliation{Max Planck Institute for Solid State Research, D-70569 Stuttgart, Germany}
\author{M. Jansen}
\affiliation{Max Planck Institute for Solid State Research, D-70569 Stuttgart, Germany}
\date{\today}

\begin{abstract}
The crystal structure and magnetic ordering pattern of PdAs$_{2}$O$_{6}$
were investigated by neutron powder diffraction. While the magnetic
structure of PdAs$_{2}$O$_{6}$ is identical to the one of its isostructural $%
3d$-homologue NiAs$_{2}$O$_{6}$, its N\'{e}el temperature (140 K) is much
higher than the one of NiAs$_{2}$O$_{6}$ (30 K). This is surprising in view
of the long distance and indirect exchange path between the magnetic Pd$^{2+}
$ ions. Density functional calculations yield insight into the electronic
structure and the geometry of the exchange-bond network of both PdAs$_{2}$O$%
_{6}$ and NiAs$_{2}$O$_{6}$, and provide a semi-quantitative explanation of
the large amplitude difference between their primary exchange interaction
parameters.
\end{abstract}

\maketitle

\section{INTRODUCTION}

The magnetic properties of transition metal compounds with $3d$ valence
electrons have been one of the central research themes in solid-state
physics for the past three decades. In view of the interplay between
magnetism and high-temperature superconductivity, particular attention has
been focused on oxides and arsenides. Recently, the electronic structure and
ordering phenomena of transition metal compounds with $4d$ and $5d$ valence
electrons (such as ruthenates and iridates) have also captured much
attention. The electronic correlations in these materials are generally
weaker than those of their $3d$ counterparts, while the spin-orbit coupling
is stronger. The quantitative description of the influence of these
parameters on the electronic phase behavior of $d$-electron compounds is an
important topic of current research. Here we report a detailed investigation
of the magnetic properties of PdAs$_{2}$O$_{6}$, a recently synthesized \cite%
{orosel} electrically insulating compound with a magnetic lattice of Pd$%
^{2+} $ ions in the electron configuration $4d^8$. We compare our results to
the isostructural compound NiAs$_{2}$O$_{6}$, which is based on Ni$^{2+}$
ions with the same number of electrons in the $3d$-shell.

PdAs$_{2}$O$_{6}$ crystallizes in the PbSb$_{2}$O$_{6}$ structure with Pd$%
^{2+}$ and As$^{5+}$ ions segregated into different layers (Fig. 1). The
octahedral coordination of Pd$^{2+}$ in this structure is unusual, because
divalent palladium shows a strong preference for square-planar coordination,
which is associated with a diamagnetic ground state. Only a few examples of
sixfold-coordinated Pd$^{2+}$ compounds are known, including the ambient-
and high-pressure polymorphs of PdF$_{2}$ as well fluoro-palladates of
composition \textit{M}PdF$_{4}$ (\textit{M} = Ca, Cd, Hg) and CsPd$_{2}$F$%
_{5}$. \cite{mueller} These compounds are paramagnetic at high temperatures
and tend to order antiferromagnetically upon cooling. In accord with this
trend, magnetic susceptibility measurements on PdAs$_{2}$O$_{6}$ showed
paramagnetic behavior at room temperature, and an antiferromagnetic phase
transition at the N\'{e}el temperature $T_{N}\sim 150$ K. \cite{orosel} This
behavior is qualitatively analogous to the one of the isostructural 3\textit{%
d}-homologues MnAs$_{2}$O$_{6}$, CoAs$_{2}$O$_{6}$, and NiAs$_{2}$O$_{6}$,
which also show antiferromagnetic ordering with $T_{N}=13$, 20, and 30 K,
respectively. \cite{nakua} However, the much higher N\'{e}el temperature of
PdAs$_{2}$O$_{6}$ is surprising, especially because the PdO$_{6}$-octahedra
do not share vertices, edges or faces. The exchange paths connecting
neighboring Pd$^{2+}$ ions are therefore long and involve at least two
bridging oxygen sites.

In order to elucidate the microscopic origin of this surprising behavior, we
have used neutron diffraction to determine the magnetic structure of PdAs$%
_{2}$O$_{6}$, which turned out to be identical to that of the 3\textit{d}%
-homologues (Section II). This implies similar networks of exchange bonds in
both sets of compounds. We employed density-functional calculations to
obtain insights into the electronic structure of PdAs$_{2}$O$_{6}$ and NiAs$%
_{2}$O$_{6}$, and specifically into the origin of the exchange paths
(which turned out to be hopping via As dimers) and of the large amplitude
difference of the primary exchange interaction parameters (Section III). A
model based on these interactions yields an excellent description of the
magnetic susceptibilities of both compounds.

\begin{figure}[h]
\centerline{\includegraphics[width=0.8%
\columnwidth,keepaspectratio]{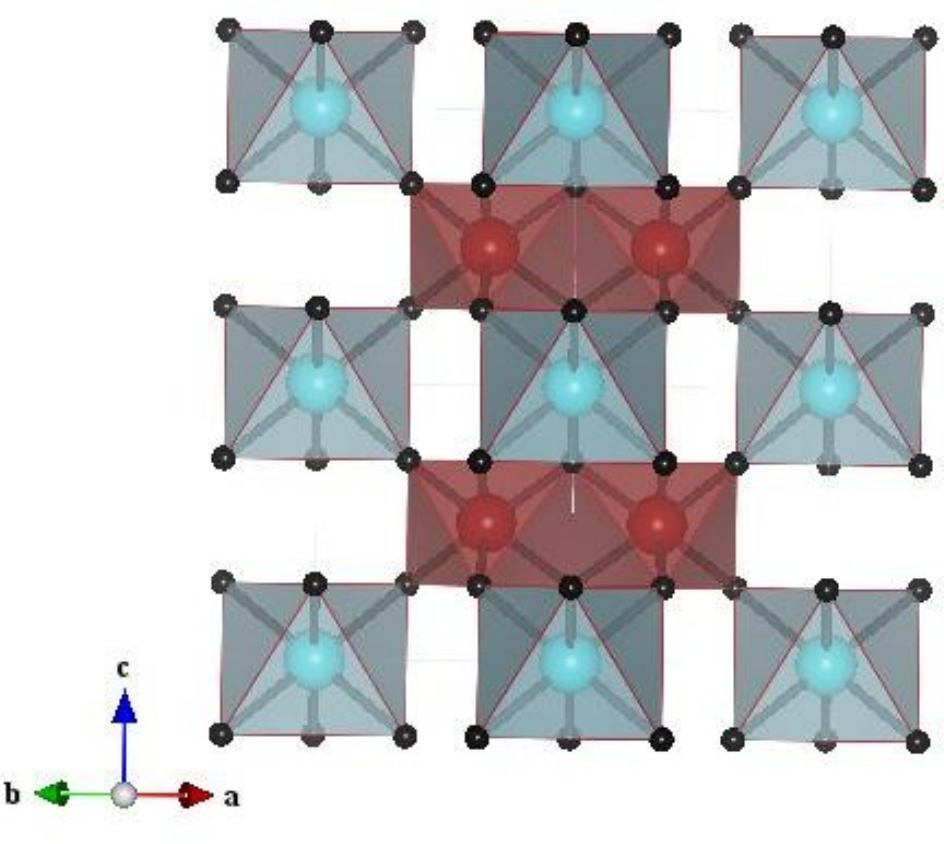}}
\caption{(Color online) Crystal structure of $A$As$_2$O$_6$ ($A$ = Pd, Ni)
showing the three-dimensional network of $AO_6$ and As$O_6$ octahedra. The
brown (dark grey) colored balls denote $A$ atoms, and the cyan (light grey)
colored balls represent As atoms. Small black colored balls at the corners
of the octahedra are O atoms.}
\label{fig1}
\end{figure}

\begin{figure}[h]
\centerline{\includegraphics[width=0.8%
\columnwidth,keepaspectratio]{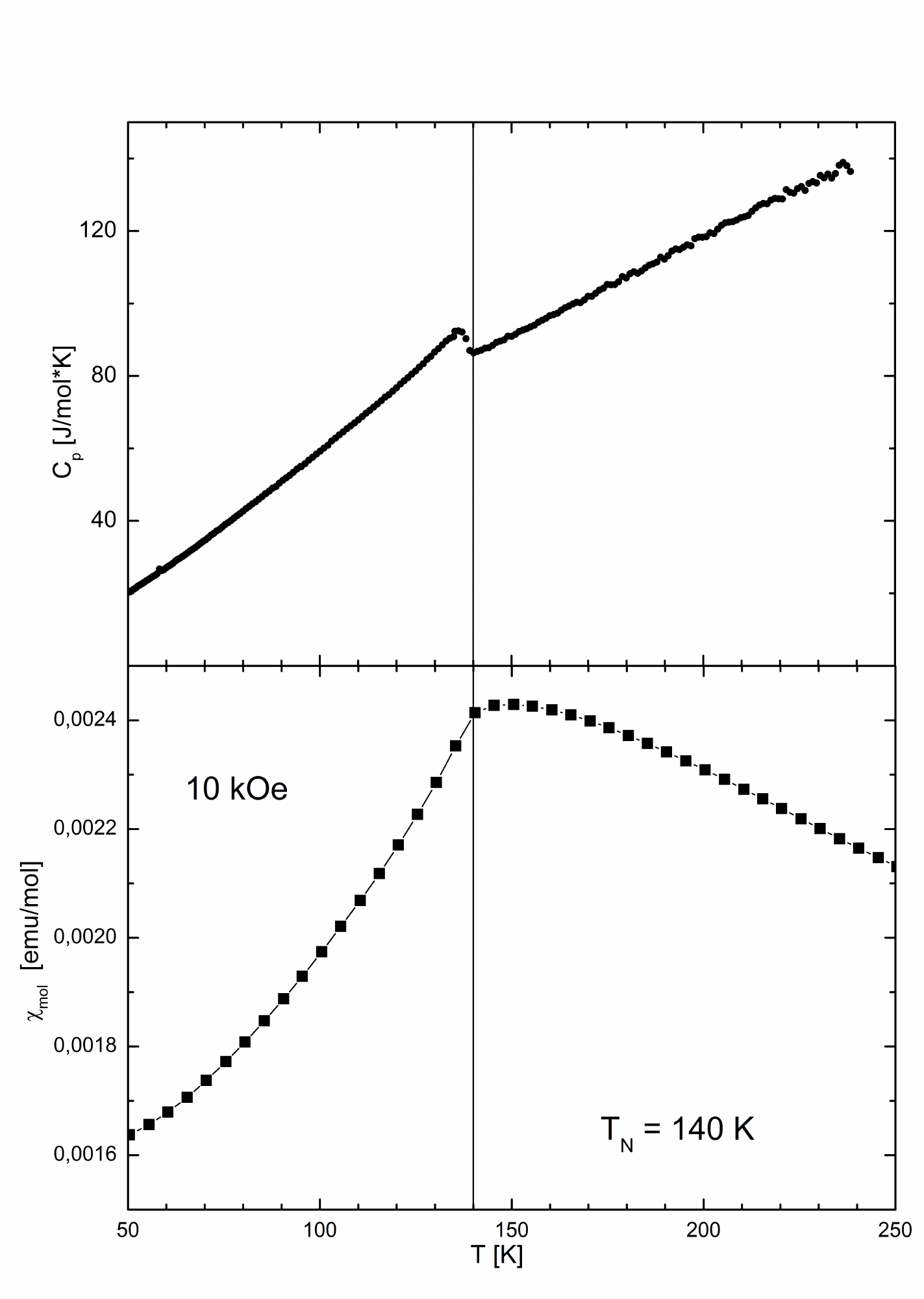}}
\caption{Temperature dependence of the specific heat and the inverse
magnetic susceptibility of PdAs$_{2}$O$_{6}$.}
\label{fig2}
\end{figure}

\section{Neutron diffraction}

\subsection{Experimental details}

A powder sample of PdAs$_{2}$O$_{6}$ of weight $\sim 3$ g was prepared using
the starting materials PdO (99.9 \% metals basis, Alfa Aesar) and As$_{2}$O$%
_{5}$ (99.9 \%, Alfa Aesar) in the molecular ratio 1 : 1.1 as described
earlier. \cite{orosel} The mixed powder was pressed into pellets and dried
in evacuated silica tubes for 12 h at 373 K. Then the evacuated silica tubes
were heated up to 973 K with a rate of 100 K/h. The hygroscopic and air
sensitive powder of PdAs$_{2}$O$_{6}$ was obtained after an annealing
process of about 100 hours. Measurements of the magnetic susceptibility and
specific heat were carried out in the temperature range between 5 and 300 K
(Fig. 2). Both quantities show anomalies indicative of antiferromagnetic
ordering of the Pd sublattice at 140 K, in agreement with prior work. \cite%
{orosel}

In order to investigate the crystal and magnetic structure of PdAs$_{2}$O$%
_{6}$, a neutron diffraction experiment was carried out at the research
reactor FRM-II in Garching. Neutron powder patterns were collected with the
instrument SPODI at 5 K and 200 K in the $2\theta$ range 4$^\circ$--160$%
^\circ$. This instrument uses a germanium monochromator (reflection 551)
selecting the neutron wavelength $\lambda = 1.5476$ \AA . The refinements of
the crystal and magnetic structure were carried out with the program
FullProf. \cite{rodriguez} We used the nuclear scattering lengths \textit{b}%
(Pd) = 5.91 fm, \textit{b}(As) = 6.58 fm and \textit{b}(O) = 5.805 fm. \cite%
{sears} The magnetic form factors of the magnetic ions were taken from Ref. %
\onlinecite{brown}.

\subsection{Crystal structure of PdAs$_{2}$O$_{6}$}

The crystal structure of PdAs$_{2}$O$_{6}$ was recently refined from x-ray
powder diffraction data in the trigonal PbSb$_{2}$O$_{6}$-type structure
(space group \textit{P}$\bar 3$1\textit{m}, No. 162), where the Pd, As and O
atoms are in the Wyckoff positions 1\textit{a}(0,0,0), 2\textit{d}($\frac{1}{%
3}$,$\frac{2}{3}$,$\frac{1}{2}$) and 6\textit{k}(\textit{x},0,\textit{z}),
respectively. \cite{orosel} The same space group was found earlier for the
compounds MnAs$_{2}$O$_{6}$, CoAs$_{2}$O$_{6}$ and NiAs$_{2}$O$_{6}$
containing 3\textit{d}-metal ions. \cite{nakua} From our neutron powder
diffraction data taken at the lower temperatures 5 and 200 K (Fig. 3) the
trigonal space group \textit{P}$\bar 3$1\textit{m} was confirmed. For the
Rietveld refinements we used data in the extended $2\theta$ range from 4$%
^\circ$ up to 146$^\circ$. A total of 14 parameters was refined: an overall
scale factor, five profile function parameters, the zero point, two lattice
constants, the positional parameters \textit{x} and \textit{z} of the oxygen
atom as well as three isotropic thermal parameters. The powder sample
contained small amounts of the binary oxide PdO, which crystallizes in the
tetragonal space group $P4_{2}/mmc$. \cite{moore} Therefore the overall
scale factor of PdO was additionally allowed to vary during the refinements.

\begin{figure*}[t]
\centerline{\includegraphics[width=1.5\columnwidth]{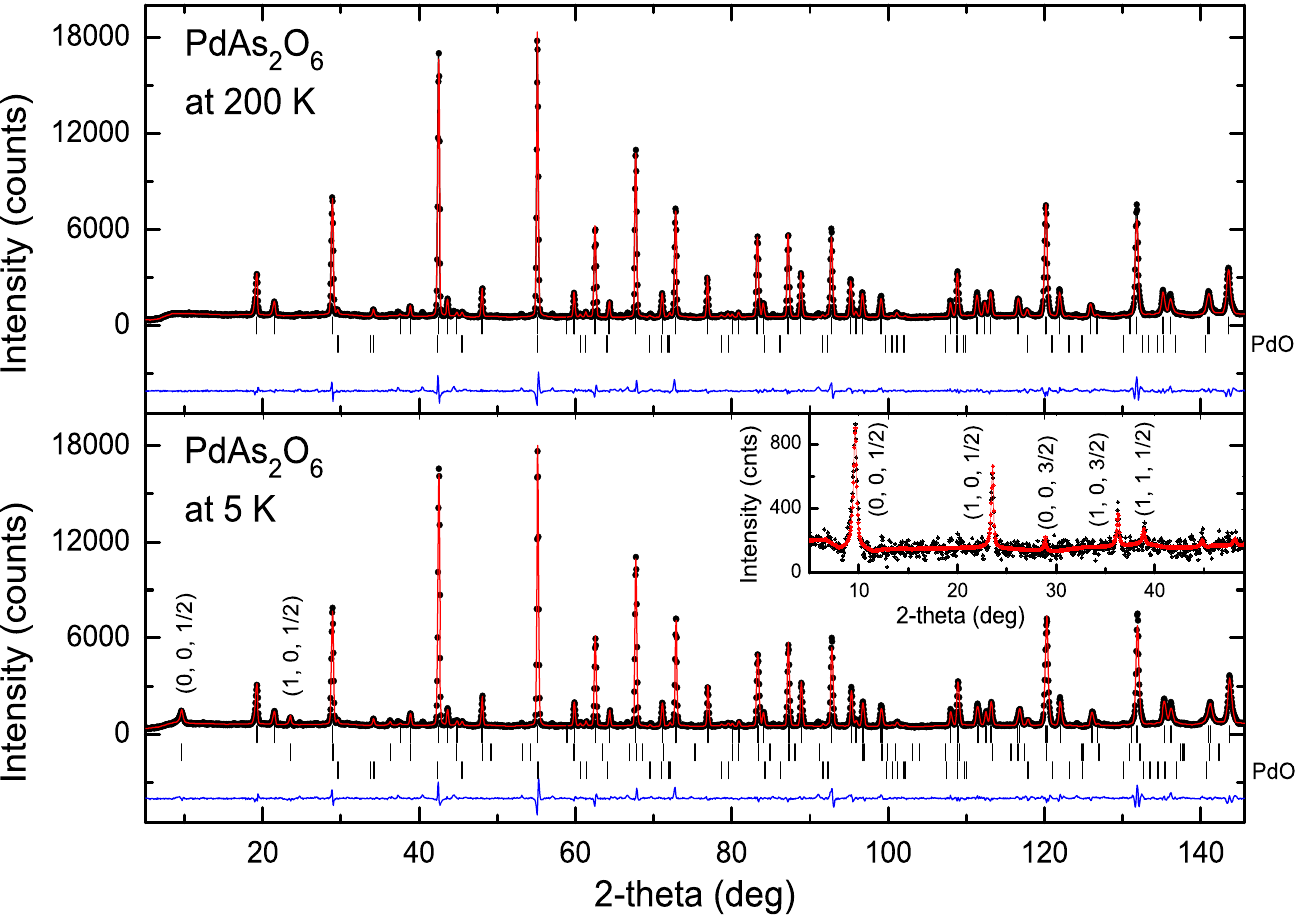}}
\caption{(Color online) Neutron powder diffraction data of PdAs$_{2}$O$_{6}$
collected at 5 and 200 K. The crystal structure was refined in the hexagonal
space group $P\overline 3 1m$. The calculated patterns (red lines) are
compared to the observed ones (black circles). In the lower part of each
diagram the difference pattern (blue) as well as the positions of the
nuclear reflections of PdAs$_{2}$O$_{6}$ and the impurity phase PdO (black
bars) are shown. For the powder pattern collected at 5 K, the positions of
the magnetic reflections of PdAs$_{2}$O$_{6}$ are also shown. The inset
shows the magnetic Bragg reflections of PdAs$_{2}$O$_{6}$, obtained from the
difference between the diffraction patterns at 5 and 200 K.}
\label{fig3}
\end{figure*}

In Table I the results of the refinements are compared with those of the
x-ray study carried out earlier at room temperature. \cite{orosel} Here it
can be seen that the positional parameters of the oxygen atoms determined at
5 and 200 K are in good agreement, indicating that the structural changes
between the magnetically ordered and the paramagnetic states are weak. Only
a slight reduction of 0.0036 \AA\ (about 6 $\sigma$) could be observed for
the Pd-O-bond length in the PdO$_{6}$-octahedra. In contrast, the distances
between the As and O-atoms are practically unchanged (Table I). The value
\textit{d}(As-O) = 1.8281(6) \AA {} (at 200 K) found for PdAs$_{2}$O$_{6}$
is in very good agreement with the values of other arsenates containing 3%
\textit{d}-metal ions: \textit{d}(As-O) = 1.827(4) \AA {} (NiAs$_{2}$O$_{6}$%
), \textit{d}(As-O) = 1.830(3) \AA {}, (CoAs$_{2}$O$_{6}$), \textit{d}(As-O)
= 1.826(2) \AA {} (MnAs$_{2}$O$_{6}$), and \textit{d}(As-O) = 1.826(1) \AA %
{} (CdAs$_{2}$O$_{6}$). \cite{nakua,weil} All of these values are in
agreement with \textit{d}(As-O) = 1.82 \AA {} calculated for an AsO$_{6}$%
-octahedron given by Shannon. \cite{shannon} In Table I it can be seen that
the structural parameters obtained at 200 and 300 K show relatively large
discrepancies, despite the fact that both data sets were collected in the
paramagnetic phase. This can be ascribed to the larger scattering power of
the O-atoms in neutron diffraction, with the consequence that the
O-positions can be determined more reliably. Furthermore, the shortest
oxygen contact \textit{d}(O-O) = 2.308(3) \AA {} was found to be implausibly
short in the x-ray study. \cite{orosel} From our neutron diffraction study
we found the larger values \textit{d}(O-O) = 2.3757(12) \AA {} at 5 K and
\textit{d}(O-O) = 2.3726(12) \AA {} at 200 K, respectively. The value
\textit{d}(O-O) = 2.410(3) \AA {} found for NiAs$_{2}$O$_{6}$ is slightly
larger, \cite{nakua} while the As-O-bond lengths \textit{d}(As-O) = 1.827(4)
\AA {} (NiAs$_{2}$O$_{6}$) and \textit{d}(As-O) = 1.8281(6) \AA {} (PdAs$_{2}
$O$_{6}$) are practically the same in both compounds. The cell volume of the
Ni compound ($V = 86.97(3)$ \AA $^{3}$) is much smaller than the one of the
Pd compound ($V = 93.698(2)$ \AA $^{3}$). This is due to fact that the ionic
radius of Pd$^{2+}$ is larger than that of Ni$^{2+}$. In order to keep the
As-O-bond lengths almost constant in the AsO$_{6}$-octahedra, the bond angle
$\angle$(O-As-O) increases from 169.91(3)$^\circ$ in PdAs$_{2}$O$_{6}$) to
173.23(12)$^\circ$ in NiAs$_{2}$O$_{6}$.

\begin{table}[h]
\begin{tabular}{|l|l|l|l|}
\hline
& 5 K & 200 K & 290 K \\ \hline
a [\AA ] & 4.81700(4) & 4.81837(5) & 4.8196(1) \\
c [\AA ] & 4.65618(6) & 4.66014(7) & 4.6646(1) \\
V [\AA $^3$] & 93.565(2) & 93.698(2) & 93.835(3) \\
x(O) & 0.37187(15) & 0.37230(16) & 0.3695(7) \\
z(O) & 0.28203(18) & 0.28236(18) & 0.2926(5) \\
\textit{B}(Pd) [\AA $^2$] & 0.46(3) & 0.62(3) & 0.80(3) \\
\textit{B}(As) [\AA $^2$] & 0.49(2) & 0.54(2) & 0.88(3) \\
\textit{B}(O) [\AA $^2$] & 0.62(2) & 0.74(2) & 0.55(6) \\
\textit{d}(Pd-O) [\AA ] & 2.2211(6) (6$\times$) & 2.2247(6) (6$\times$) &
2.2437(20) (6$\times$) \\
$\angle$(O-Pd-O) [$^\circ$] & 180 (3$\times$) & 180 (3$\times$) & 180 (3$%
\times$) \\
& 88.61(2) (6$\times$) & 88.58(2) (6$\times$) & 86.84(10) (6$\times$) \\
& 91.39(2) (6$\times$) & 91.42(2) (6$\times$) & 93.16(10) (6$\times$) \\
\textit{d}(As-O) [\AA ] & 1.8288(6) (6$\times$) & 1.8281(6) (6$\times$) &
1.8076(23) (6$\times$) \\
$\angle$(O-As-O) [$^\circ$] & 169.91(3) (3$\times$) & 169.79(3) (3$\times$)
& 170.42(13) (3$\times$) \\
& 81.01(3) (3$\times$) & 80.92(4) (3$\times$) & 79.34(15) (3$\times$) \\
& 92.18(3) (3$\times$) & 92.19(3) (3$\times$) & 93.34(15) (3$\times$) \\
& 95.49(2) (6$\times$) & 95.57(2) (6$\times$) & 94.03(10) (6$\times$) \\
\textit{d}(O-O)$_{min}$ [\AA ] & 2.3757(12) & 2.3726(12) & 2.308(3) \\ \hline
$R_N$ & 0.0241 & 0.0298 & 0.0859$^\ast$ \\ \hline
\end{tabular}%
\caption{Results of Rietveld refinements of the neutron powder diffraction
data ($\protect\lambda = 1.5476$ \AA ) for the nuclear structure of PdAs$_{2}
$O$_{6}$ at 5 and 200 K. The lattice constants, positional and isotropic
thermal parameters as well as the bond distances and angles within the AsO$%
_{6}$- and PdO$_{6}$-units are compared with the values obtained earlier at
room temperature from x-ray powder diffraction data ($\protect\lambda =
0.7093$ \AA ). \protect\cite{orosel} The residual \textit{R}$_{N}$ of the
refinement of the crystal structure is defined as $R_N = \sum ||(F_{obs}| -
|F_{calc}||/|(F_{obs}|$. The residual for the room temperature structure
(marked by $\ast$) was calculated with intensities rather than structure
factors. \protect\cite{orosel}}
\label{table1}
\end{table}

\subsection{Magnetic structure of PdAs$_{2}$O$_{6}$}

The neutron powder data recorded at 5 K show additional Bragg reflections
that can be ascribed to the antiferromagnetic order of the Pd sublattice.
The two prominent ones at $2\theta = 9.5^\circ$ and $2\theta = 23.5^\circ$
can be indexed as (0, 0, $\frac{1}{2}$)$_{M}$ and (0, 1, $\frac{1}{2}$)$_{M}$%
, respectively. This suggests that the magnetic cell has a doubled \textit{c}%
-axis with a propagation vector \textbf{\textit{k}} = (0, 0, $\frac{1}{2}$).
All further magnetic reflections were assigned indices according to (\textit{%
hk$\ell$})$_{M}$ = (\textit{hk$\ell$})$_{N}$ $\pm$ \textbf{\textit{k}},
where M and N designate the magnetic and nuclear reflections. The same type
of magnetic ordering was observed earlier for the isotypic divalent
transition metal arsenates CoAs$_{2}$O$_{6}$ and NiAs$_{2}$O$_{6}$. \cite%
{nakua} The presence of the strong magnetic reflection (0, 0, $\frac{1}{2}$)$%
_{M}$ indicates that the magnetic moments of the Pd-atoms are aligned
ferromagnetically within the hexagonal \textit{ab}-plane. Due to
antiferromagnetic exchange interactions between the palladium moments the
ferromagnetic layers form the sequence + - + - along the \textit{c}-axis
(Fig. 4). With this model the magnetic structure of PdAs$_{2}$O$_{6}$ could
be successfully refined using the magnetic reflections observed in the $%
2\theta$-range up to 50$^\circ$. It has to be noted that the moment
direction within the hexagonal plane cannot be determined from the
refinements. The wave vector \textbf{\textit{k}} = (0, 0, $\frac{1}{2}$)
keeps the full symmetry of the group \textbf{\textit{G$_{k}$}} = \textbf{%
\textit{G}} according to magnetic group theory, and it defines the magnetic
translation lattice. \cite{rossat} The existence of three magnetic domains
in the hexagonal basis plane prohibits an unambiguous determination of the
moment direction.

\begin{table}[h]
\begin{tabular}{|l|l|l|l|l|}
\hline
& $I_{obs}$ & $I_{calc}$ (Ni$^{2+}$) & $I_{calc}$ (Pd$^+$) & $2\theta
(^\circ)$ \\ \hline
(0, 0, $\frac{1}{2}$)$_M$ & 518 & 445 & 503 & 9.5 \\
(1, 0, $\frac{1}{2}$)$_M$ & 198 & 205 & 178 & 23.5 \\
(0, 0, $1\frac{1}{2}$)$_M$ & 9 & 34 & 25 & 28.8 \\
(1, 0, $1\frac{1}{2}$)$_M$ & 88 & 87 & 50 & 36.2 \\
(1, 1, $\frac{1}{2}$)$_M$ & 42 & 45 & 23 & 38.8 \\
(2, 0, $\frac{1}{2}$)$_M$ & 30 & 28 & 11 & 44.8 \\
(1, 1, $1\frac{1}{2}$)$_M$ & 43 & 28 & 9 & 48.0 \\
(0, 0, $2\frac{1}{2}$)$_M$ & 2 & 6 & 2 & 49.1 \\ \hline
$\mu_{exp}$ (Pd$^{2+}$) ($\mu_B$) &  & 1.87(3) & 2.04(3) &  \\ \hline
$R_M$ &  & 0.141 & 0.173 &  \\ \hline
\end{tabular}%
\caption{Observed and calculated intensities of the magnetic reflections of
PdAs$_{2}$O$_{6}$ as obtained from Rietveld refinements using the magnetic
form factors of Ni$^{2+}$ and Pd$^+$. \protect\cite{brown} The residual $R_M$
is defined as $R_M = \sum ||(I_{obs}| - |I_{calc}||/|(I_{obs}|$.}
\label{table2}
\end{table}


In order to improve the refinement of the magnetic structure, we used the
purely magnetic intensities obtained from the difference between the data
sets collected at 5 and 200 K (inset of Fig. 3). Since the magnetic form
factor of Pd$^{2+}$ is not available, we first used the one of the Pd$^{+}$
ion, \cite{brown} but Table II shows that the calculated intensities
decrease much more strongly with increasing $2\theta$ than the observed
ones. A considerably better fit was obtained with the form factor of Ni$%
^{2+} $, which also shows a $d^8$-configuration. The relatively large
residual $R_M = 0.141$ [defined as $R_M = \sum ||(I_{obs}| -
|I_{calc}||/|(I_{obs}|$] reflects the fact that the low intensities of the magnetic
reflections at high $2\theta$-values could not be measured with good
accuracy, in combination with systematic errors arising from the difference
between the form factors of Pd$^{2+}$ and Ni$^{2+}$.

The sublattice magnetization resulting from the refinement is $\mu = 1.92(4)
\mu_B$ per Pd site, similar to the value $\mu = 2.11(3) \mu_B$ reported for
NiAs$_{2}$O$_{6}$ (Ref. \onlinecite{nakua}). While the ordered moment is
consistent with the spin-only moment expected for a $d^8$ configuration, a
fit to the magnetic susceptibility for $T > T_N$ yields a $g$-factor larger
than 2, which is indicative of an orbital contribution to the Pd moment (see
Section III.C below). The difference may in part be due to zero-point
fluctuations of the magnetic moment, which reduce the ordered moment of the
spin-1 system in the binary oxide NiO by $\sim 8$ \%. \cite{hutchings} The
zero-point reduction is possibly larger in PdAs$_{2}$O$_{6}$ because of the
low-dimensional exchange-bond network (see Section III.C). While these
considerations suggest a small but nonzero orbital contribution to the Pd
moment, measurements of the $g$-factor anisotropy by single-crystal neutron
diffraction and/or electron spin resonance will be required to quantify this
contribution.

\begin{figure}[h]
\centerline{\includegraphics[width=0.8%
\columnwidth,keepaspectratio]{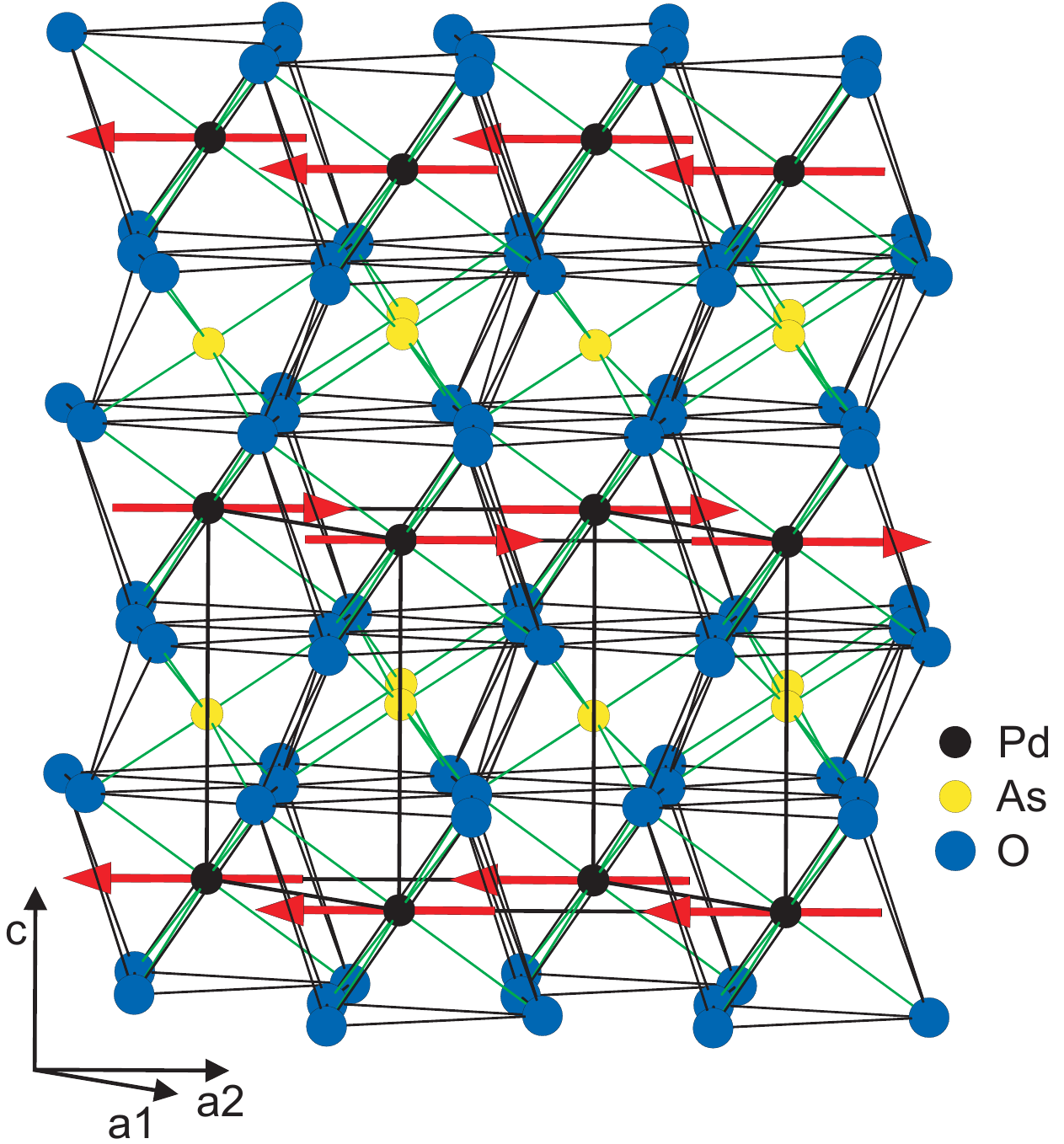}}
\caption{(Color online) Magnetic structure of PdAs$_{2}$O$_{6}$. Shown are
the isolated PdO$_{6}$-octahedra between the AsO$_{6}$-layers. The magnetic
moments of the palladium atoms are aligned ferromagnetically within the
hexagonal \textit{ab}-plane. Along the \textit{c}-direction the moments are
coupled in an antiparallel fashion, forming the sequence + - + -.}
\label{fig4}
\end{figure}

\section{Density functional calculations}

\subsection{LDA band structure}

Figure 5 shows the electronic band structure and density of states in the
paramagnetic local density-functional approximation (LDA\cite{LDA1}) for NiAs%
$_{2}$O$_{6}$ in solid lines and for PdAs$_{2}$O$_{6}$ in dashed lines. The
self-consistent calculations were performed with the linear muffin-tin
orbital (LMTO) method~\cite{LMTO1} using $8\times 8\times 8$ $\mathbf{k}$%
-points in the Brillouin zone.

The Fermi level falls in the middle of the two narrow transition-metal $%
\left( A\right) $ $e_{g}$-bands. These are split above the three narrow $%
t_{2g}$-bands, because the $pd\sigma $-hopping is $\sim \sqrt{3}$ times
stronger than the $pd\pi $-hopping to the O $p$ orbitals at the corners of
the $A\mathrm{O}_{6}$-octahedron. This is in accord with the labeling $%
A^{2+}.$ Similarly, in accord with the labeling O$^{2-}$ and As$^{5+},$ the
oxygen $2p$-like bands are below and the As $s$- and As $p$-like bands are above
the Fermi level. However, in terms of atomically localized orbitals (LMTOs)
rather than Wannier orbitals, the bands denoted as As $s$ in the figure have
$\sim $40\% anti-bonding O $p$ character, as well as some $A$ $s$ and O $s$
characters. Correspondingly, around $-$12 eV (below the frame of the figure)
there are two bands with As $s$ and O $sp$ bonding characters in about equal
amounts.

With the bands lined up at the Fermi level, the $t_{2g}$ and $p$ bands lie
lower in the Pd than in the Ni compound. This is because $4d$ orbitals have
a larger extent and therefore larger hopping integrals to O $p$ than do $3d$
orbitals. For the same reason, the $e_{g}$ band is about 1.4 times wider in
the Pd than in the Ni compound.

Inclusion of the Coulomb interaction beyond the LDA splits the $e_{g}$ bands
and leads to insulating solutions, as we have checked through LDA+U~\cite%
{LDA+U} calculations. For the present purpose of calculating and
understanding the magnetic properties, it is more convenient to start from
the localized description and treat the hopping, $t$, to order $t^{2}/U.$

\begin{figure}[t]
\centerline{\includegraphics[width=\columnwidth]{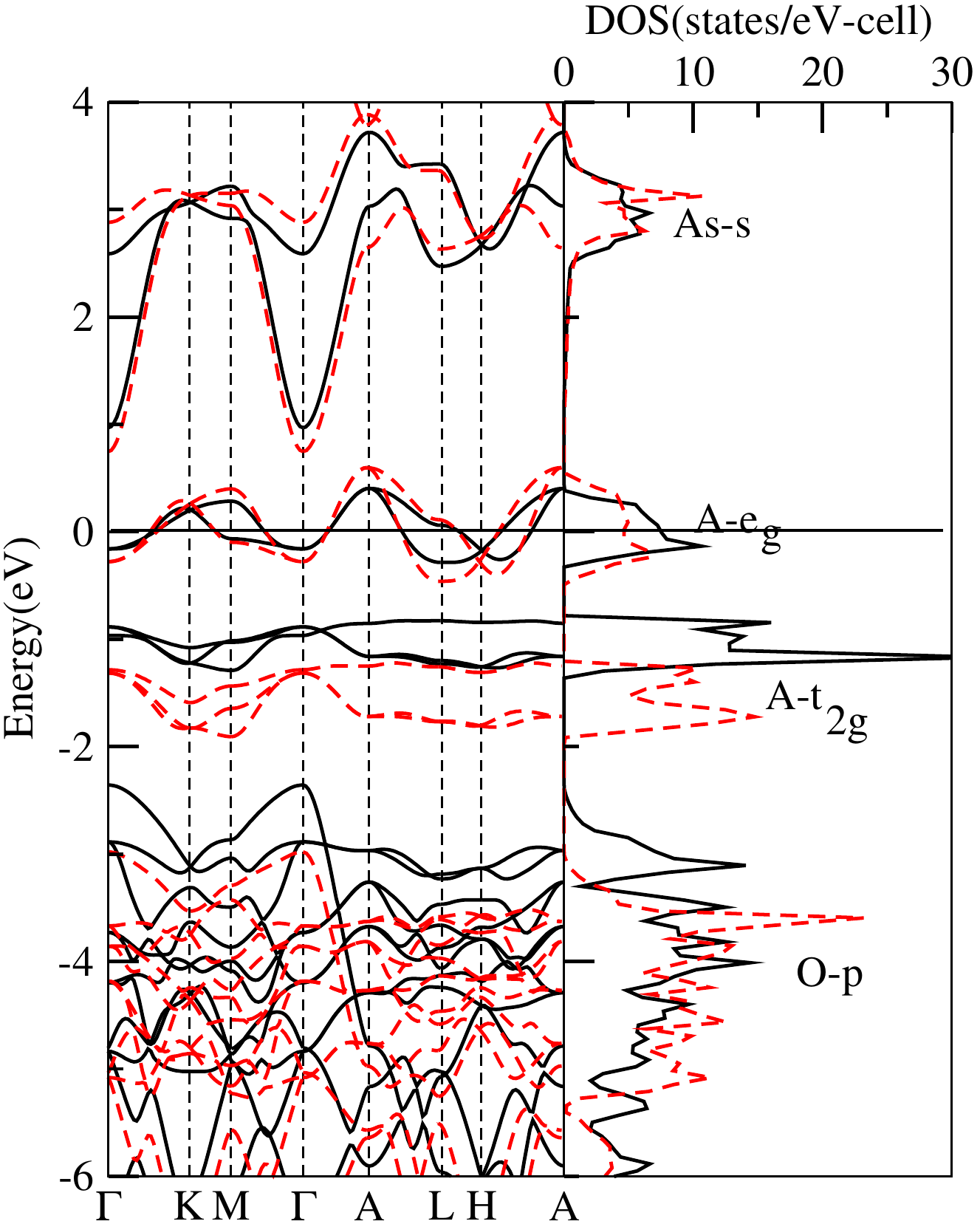}}
\caption{(Color online) LDA band structure (left panel) and density of
states (right panel) for NiAs$_{2}$O$_{6}$ (solid lines) and for PdAs$_{2}$O$%
_{6}$ (dashed lines). The two band structures have their Fermi levels (0 eV)
lined up. The bands are plotted along the high-symmetry directions of the
hexagonal Brillouin zone and the densities of states have their dominating
characters labeled.}
\label{fig5}
\end{figure}

\subsection{Pd$\,e_{g}$ Wannier orbitals, low-energy tight-binding
Hamiltonian and magnetic interactions}

We therefore construct a low-energy Hubbard Hamiltonian. Since the LDA $e_{g}
$ band is narrow and well separated from all other bands, we can limit the
one-electron Hilbert space to that of the two $A$-centered Wannier orbitals,
$3z^{2}-r^{2}$ and $x^{2}-y^{2},$ describing this band. When using the $N$%
MTO downfolding method, we thus kept the $A$ $e_{g}$ degrees of freedom as
active, and downfolded the rest. The Wannier orbitals are finally obtained
by symmetrically orthonormalizing the downfolded, $N$th-order muffin-tin
orbitals ($N$MTOs~\cite{NMTO}). In the representation, $H=\sum
t_{im,jm^{\prime }}(\hat{c}_{im}^{\dagger }\hat{c}_{jm^{\prime }}+h.c.),$ of
the corresponding one-electron part of the Hamiltonian, $t_{im,jm^{\prime }}$
is the hopping integral from orbital $m$ on site $i$ to orbital $m^{\prime }$
on site $j.$

\begin{figure*}[t]
\includegraphics[width=1.8\columnwidth]{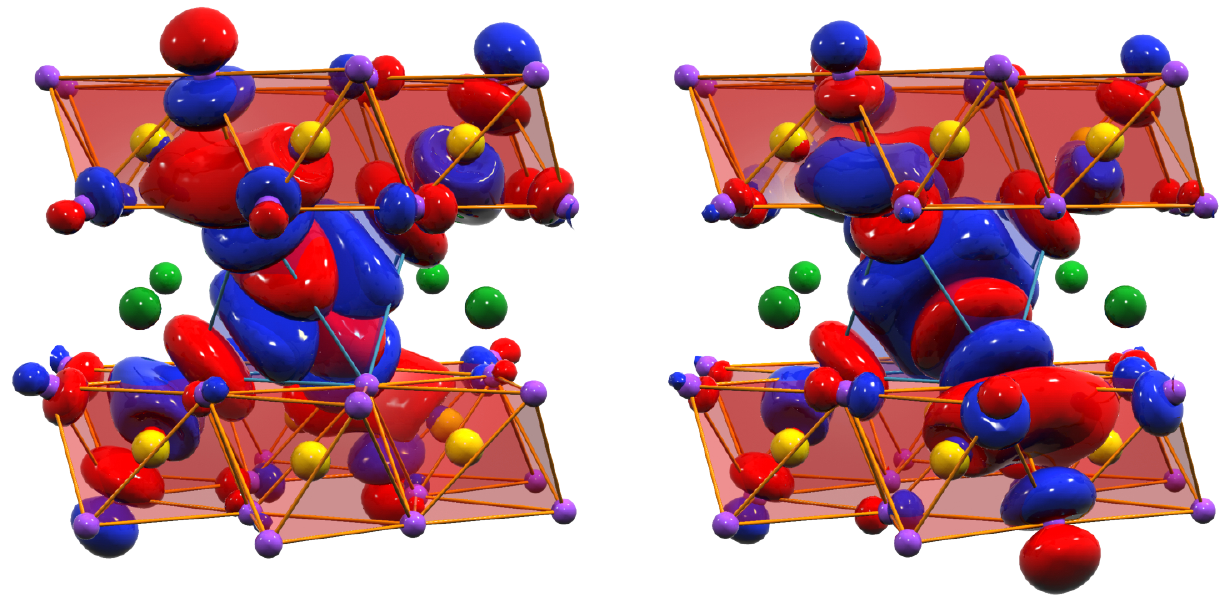}
\caption{(Color online) The $x^{2}-y^{2}$ (left panel) and $3z^{2}-r^{2}$
(right panel) Wannier orbitals which span the LDA Pd $e_{g}$ bands of PdAs$%
_{2}$O$_{6}$. Plotted are orbital shapes (constant-amplitude surfaces) with
lobes of opposite signs colored red (dark) and blue (light). For clarity,
the central PdO$_{6}$ octahedron is given a blue skin and all AsO$_{6}$
octahedra a red skin. Pd atoms are green, As atoms yellow, and oxygens
violet.}
\label{fig6}
\end{figure*}


Figure \ref{fig6} shows the $x^{2}-y^{2}$ and $3z^{2}-r^{2}$ Wannier
orbitals for PdAs$_{2}$O$_{6}$ as $\chi \left( \mathbf{r}\right) =\pm const.$
surfaces with positive lobes colored red and negative blue. The $z$-axis
points along the Pd-O bond in the $bc$-plane and the $x$- and $y$-axes point
to the other oxygens in the Pd-centered octahedron. In the figure, the
latter has been given a blue, transparent skin while all As-centered
octahedra have been given a red, transparent skin. Pd atoms are green, As
atoms yellow, and oxygens violet. Such a Pd-centered Wannier orbital has $%
x^{2}-y^{2}$ or $3z^{2}-r^{2}$ character locally, and strong $pd\sigma $
antibonding character on the neighboring oxygens. The unusual feature here is
that the \emph{back}-lobes of the strongest O $p$ characters (the four $p_{x}
$ and $p_{y}$ tails of the $x^{2}-y^{2}$ Wannier orbital and the two $p_{z}$
tails of the $3z^{2}-r^{2}$ Wannier orbital) bond to the $sp$ characters on
the closest \emph{pair of As atoms} and, from there, antibond to the closest
oxygen, which now belongs to a neighboring PdO$_{6}$ octahedron. As an
example: The red lobe of the $x^{2}-y^{2}$ orbital sticking up and out
towards the reader, antibonds with the blue lobe of near $p_{x}$ orbital,
whose red back-lobe merges together (bonds) with the red, two-center As $sp$
bond, giving rise to a ``red sausage''. The latter finally antibonds with the
O $p$ orbital which points upwards towards a near Pd belonging to the upper
Pd-sheet (not shown in the figure). Similarly for the $3z^{2}-r^{2}$ orbital
on the right-hand side of the figure: Its red lobe, sticking down and out,
antibonds with the blue lobe of the near $p_{z}$ orbital, which merges with
the As two-center bond into a red sausage. The latter finally induces strong
$p_{z}$-like character which points down and out, towards a Pd atom in the
lower Pd sheet.

\begin{figure}[tbp]
\begin{center}
\includegraphics[width=0.8\columnwidth,keepaspectratio]{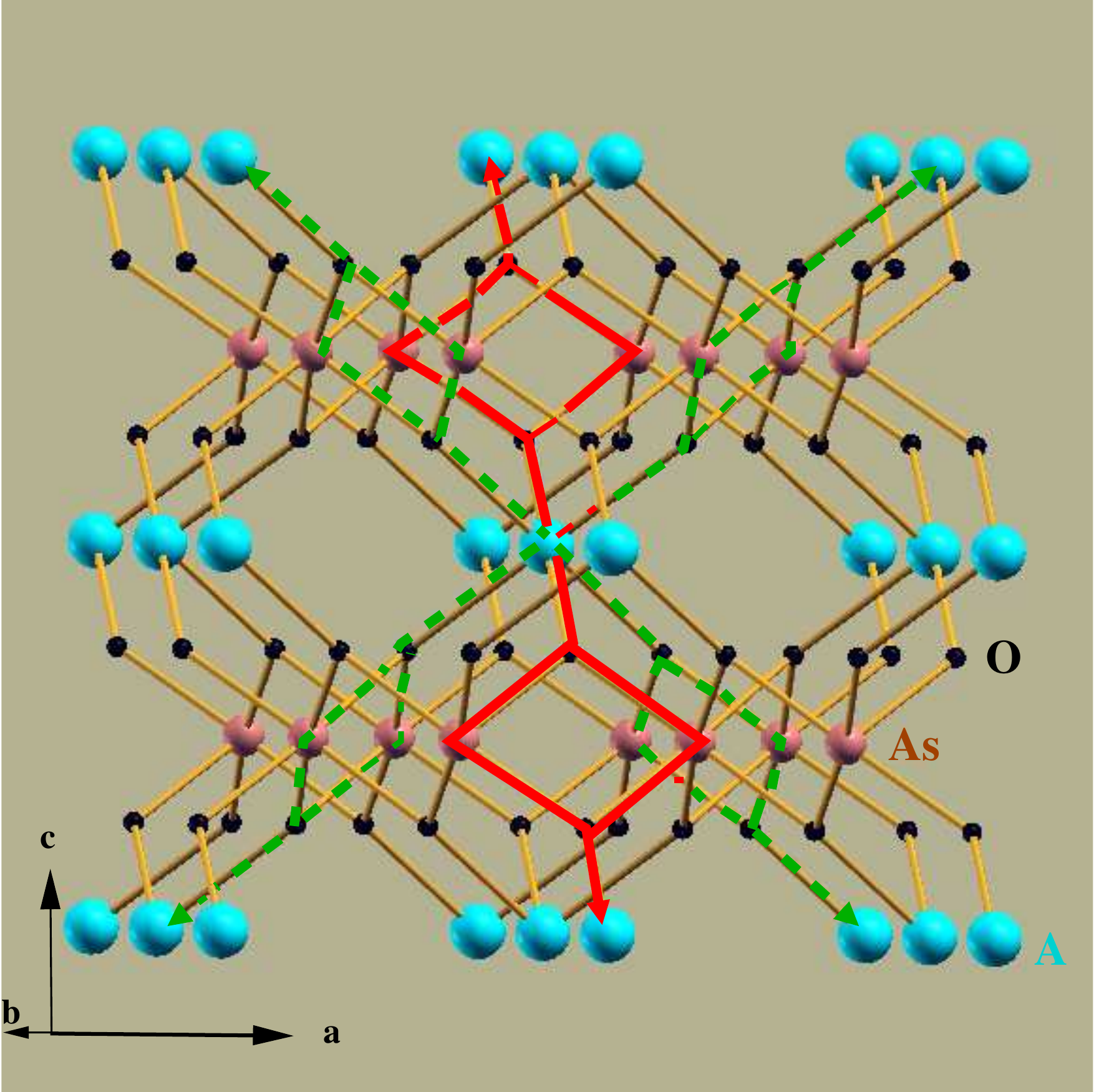}
\end{center}
\caption{(Color online) Paths of the dominant hoppings. Dashed green:
between $x^{2}-y^{2}$ orbitals (see left-hand side of Fig. 6) on 3rd-nearest
$A$ neighbors. Solid red: between $3z^{2}-r^{2}$ orbitals (see right-hand
side of Fig. 6) on 3rd-nearest $A$ neighbors. The colors of the atoms are as
in Fig. 1.}
\end{figure}

These hopping paths, shown schematically in Fig. 7, are to
3rd-nearest Pd neighbors, but only to six out of the twelve, namely to those
at $\pm \left( 0,1,1\right) ,$ $\pm \left( \frac{\sqrt{3}}{2},\frac{1}{2}%
,-1\right) ,$ and $\pm \left( \frac{\sqrt{3}}{2},-\frac{1}{2},1\right).$
The $e_{g}$ integrals for hopping to the 3rd-nearest neighbors at $\pm
(0,1,-1),$ $\pm (\frac{\sqrt{3}}{2},\frac{1}{2},1),$ and $\pm \left( -\frac{%
\sqrt{3}}{2},\frac{1}{2},1\right) $, which have no bridging oxygen and As pairs, are negligible. The calculated hopping
integrals exceeding 10 meV are given in Table III. We see that those to
3rd-nearest neighbors dominate those to 2nd- and 1st-nearest neighbors.

The Ni $3d\left( e_{g}\right) $ Wannier orbitals in NiAs$_{2}$O$_{6}$ are
similar to the Pd $4d\left( e_{g}\right) $ Wannier orbitals in PdAs$_{2}$O$%
_{6}$, except that they are more localized. This is consistent with the 1.4
times smaller $3d\left( e_{g}\right) $-band width. Concomitantly, the
hopping integrals for NiAs$_{2}$O$_{6}$ listed in parentheses in Table III
are about 1.4 times larger than those for PdAs$_{2}$O$_{6}$.

\begin{table*}[t]
\begin{tabular}{|c||c|c|c||c|c|c|c|}
\hline
Vector from $m$ to $m^{\prime } \rightarrow$ &   $\pm \left( 0,1,0\right) $ & $\pm
\left( \frac{\sqrt{3}}{2},\frac{1}{2},0\right) $ & $\pm \left( \frac{\sqrt{3}%
}{2},-\frac{1}{2},0\right) $ &  $\pm \left( 0,1,1\right) $ & $\pm \left(
\frac{\sqrt{3}}{2},\frac{1}{2},-1\right) $ & $\pm \left( \frac{\sqrt{3}}{2},-%
\frac{1}{2},1\right) $ \\
$m,m^{\prime }\downarrow $ &  &  &  &  &  &  \\ \hline
$3z^{2}-r^{2},\,3z^{2}-r^{2}$ &   $33\;(24)$ & .\ $\left( .\right) $ & .\ $%
\left( .\right) $ &  $-139\;(-99)$ & $-38\;(-27)$ & $-38\;(-27)$ \\ \hline
$x^{2}-y^{2},\,x^{2}-y^{2}$ &  $-12\;(-12)$ & $22\;(15)$ & $22\;(15)$ &
.\ $\left( .\right) $ & $-106\;(-76)$ & $-106\;(-76)$ \\ \hline
$3z^{2}-r^{2},\,x^{2}-y^{2}$ &  .\ $\left( .\right) $ & $-19\;(-16)$ & $%
19\;(16)$ & .\ $\left( .\right) $ & $58\;(42)$ & $-58\;(-42)$ \\
\hline
\end{tabular}
\caption{Hopping integrals between Wannier $e_{g}$ orbitals centered on Pd
(or Ni). Hoppings are in meV and listed are those exceeding 10 meV. Only
hoppings between 2nd-nearest neighbors, $d_{2}$= 4.82 (4.76) \AA , and
between 3rd-nearest neighbors, $d_{3}$ = 6.71 ( 6.51) \AA , are significant.}
\label{table3}
\end{table*}

The magnetic exchange interaction, $J$, can be expressed in general as a sum
of antiferromagnetic and ferromagnetic contributions. In the limit of large
Coulomb correlation, typically valid for late transition metal elements like
Pd or Ni, the antiferromagnetic contribution is related to the hopping
integral $t$ by the second-order perturbation relation $J$ $\sim \frac{t^{2}%
}{U}$, where $U$ is the effective on-site Coulomb repulsion defined for the
Wannier orbitals. Considering the hopping integrals in Table III, the
contribution from the $t^{2}$ term in the magnetic exchange gives rise to a
factor two difference between the Pd and Ni compounds. The estimate of $U$,
unlike that of the hopping integral $t$, is a rather delicate issue. In
principle, one should compute $U$ for the Wannier orbitals shown in Figure
7. However lacking a prescription to do so, we computed the $U$ values
corresponding to Ni-$d$ and Pd-$d$ partial waves, which were truncated
outside the atomic spheres defined around an Ni or Pd sites. The
calculations were carried out within the framework of the constrained DFT
scheme~\cite{DFT,DFT1,DFT2}. The $U$ values calculated in this way were $%
U_{Pd}=5.7$ eV and $U_{Ni}=8.4$ eV. But since the Wannier $e_{g}$ orbitals
are far more delocalized than the truncated partial waves, and more so for
Pd than for Ni, these $U$ values should be significantly reduced, and more
so for Pd than for Ni. This could conceivably lead to the factor $%
U_{Ni}/U_{Pd}\sim 2.4$ needed to bring our theoretical $t^{2}/U$ estimate
for the N\'{e}el temperatures of the two compounds into agreement with the
measured ratio of 4.7.

\subsection{Spin model and susceptibility}

Taking into account only the dominant hopping integrals provided by the $N$%
MTO-downfolding calculation, a spin model can be defined in terms of the six
pairs of 3rd-nearest-neighbor magnetic interactions, all of size $J_{3}$,
obtained by summing over the squares of the $e_{g}$hopping integrals between
the 3rd-nearest neighbors. Based on this model, we have calculated the
magnetic susceptibility by considering the following spin-1 Hamiltonian:

\begin{eqnarray*}
H&=&J_3\sum_{k=0}^{m-1}\sum_{j=0}^{m-1}\sum_{i=0}^{m-1}
(S_{i,j,k}S_{i,j+1,k+1} + S_{i,j,k}S_{i+1,j,k+1} \\
& +& S_{i,j,k}S_{i+1,j+1,k-1})
\end{eqnarray*}

This model was solved by the quantum Monte Carlo method \cite{QMC} on a $%
10\times 10\times 10$ lattice. The primary interaction $J_{3}$, and the
effective Land\'{e} $g$-factors were obtained by fitting to the experimental
susceptibility. As shown in Fig. 8, the calculated and measured
susceptibilities agree very well. The optimal values of the $g$-factor and
the exchange parameter $J_{3}$, were found to be respectively 2.38 and 62 K
for the Pd compound, and 2.48 and 13 K for the Ni compound. We find that the
$g$-factors are larger than the spin-only value of 2, in accord with the
discussion in Section II.C above. The fit of the susceptibilities yields a
value of 5 for the ratio of the predominant magnetic exchange parameters of
PdAs$_{2}$O$_{6}$ and NiAs$_{2}$O$_{6}$, in good agreement with the ratio of
the magnetic transition temperatures.

\begin{figure}[h]
\includegraphics[width=\columnwidth,keepaspectratio]{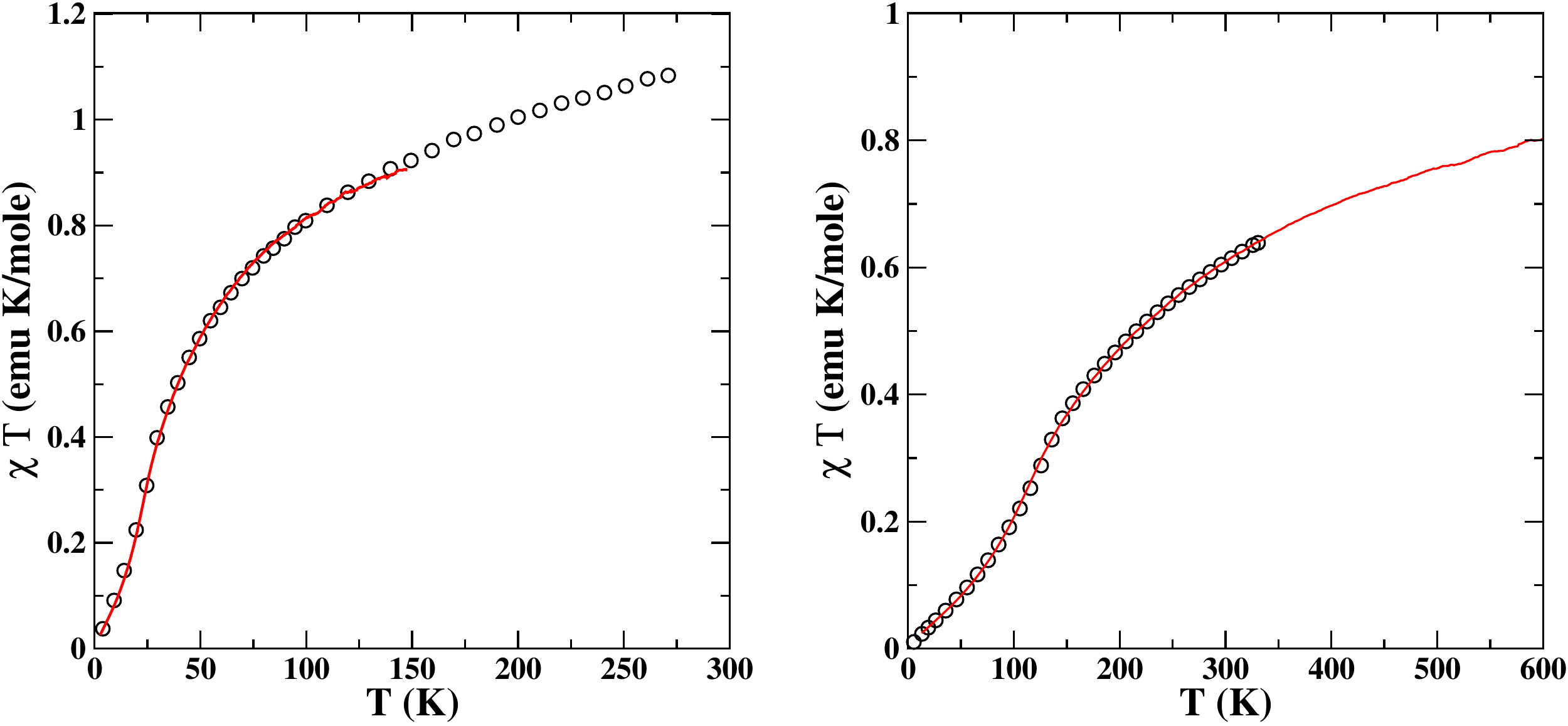}
\caption{(Color online) Temperature dependence of the magnetic susceptibility
(multiplied by temperature) for NiAs$_{2}$O$_{6}$ (left panel) and PdAs$_{2}$%
O$_{6}$ (right panel). The circles correspond to experimental data (taken
from Ref. \onlinecite{nakua} for NiAs$_{2}$O$_{6}$), and the solid lines
correspond to calculated data based on the model described in the text.}
\label{fig8}
\end{figure}

\section{Conclusions}

Using a combination of susceptibility and neutron diffraction measurements,
we have developed a comprehensive experimental description of the magnetic
properties of the newly synthesized antiferromagnet PdAs$_{2}$O$_{6}$.
Density functional theory has provided a detailed understanding of the
magnetic bond network of this compound, as well as a semi-quantitative
explanation of the large enhancement of the magnitude of its primary
exchange interaction parameters compared to its $3d$ homologue NiAs$_{2}$O$%
_{6}$. This approach may prove useful for research on other Pd compounds
including the recently discovered \cite{bruns} ferromagnet PdS$_{2}$O$_{7}$,
and for comparative studies of materials with $4d$ valence electrons and
their $3d$-electron counterparts in general.

\section{Acknowledgements}

We are grateful to the referee for constructive criticism, to M. H\"{o}lzel
and A. Senyshyn for help with the neutron diffraction measurements at the
FRM-II, and to O. Jepsen for performing the $U$ calculations. TSD and OKA
acknowledge the MPG-India partner group program and the INDO-EU research network RP7 MONAMI for support.

\end{document}